\documentclass{aastex631}

\usepackage{CJK}
\usepackage{amsmath}
\newcommand{\HI}{H\,{\textsc{I}}}
\newcommand{\HII}{H\,{\textsc{\romannumeral 2}}}

\definecolor{darkblue}{rgb}{0.0, 0.0, 0.55}

\shorttitle{PSR~B1937+21 H I absorption \& anomalous dispersion}
\shortauthors{Jiang et al.}
\graphicspath{{./}{figures/}}

\begin{document}
\begin{CJK*}{UTF8}{gbsn}
\title{H I absorption line and anomalous dispersion in the radio pulses of PSR~B1937+21}

\correspondingauthor{Jinchen Jiang, Kejia Lee, Renxin Xu}
\email{jiangjinchen@pku.edu.cn, kjlee@pku.edu.cn, r.x.xu@pku.edu.cn}

\author[0000-0002-6465-0091]{Jinchen Jiang (姜金辰)}
\affiliation{National Astronomical Observatories, Chinese Academy of Sciences, Beijing 100101, China}
\affiliation{Department of Astronomy, Peking University, Beijing 100871, China}

\author[0009-0007-3817-8188]{Shunshun Cao (曹顺顺)}
\affiliation{Department of Astronomy, Peking University, Beijing 100871, China}
\affiliation{State Key Laboratory of Nuclear Physics and Technology, Peking University, Beijing 100871, China}

\author[0000-0002-1435-0883]{Kejia Lee (李柯伽)}
\affiliation{Department of Astronomy, Peking University, Beijing 100871, China}
\affiliation{Kavli Institute for Astronomy and Astrophysics, Peking University, Beijing 100871, China}
\affiliation{National Astronomical Observatories, Chinese Academy of Sciences, Beijing 100101, China}

\author[0000-0002-9434-4773]{Bojun Wang (王铂钧)}
\affiliation{National Astronomical Observatories, Chinese Academy of Sciences, Beijing 100101, China}

\author[0000-0002-5031-8098]{Heng Xu (胥恒)}
\affiliation{National Astronomical Observatories, Chinese Academy of Sciences, Beijing 100101, China}

\author[0000-0002-3118-5963]{Siyuan Chen}
\affiliation{Shanghai Astronomical Observatory, Chinese Academy of Sciences, Shanghai 200030, China}

\author[0009-0008-1501-273X]{Yanjun Guo (郭彦君)}
\affiliation{National Astronomical Observatories, Chinese Academy of Sciences, Beijing 100101, China}
\affiliation{Key Laboratory of Radio Astronomy and Technology, Chinese Academy of Sciences, Beijing 100101, China}

\author{Peng Jiang (姜鹏)}
\affiliation{National Astronomical Observatories, Chinese Academy of Sciences, Beijing 100101, China}
\affiliation{Guizhou Radio Astronomical Observatory, Guizhou University, Guiyang 550000, China}

\author[0000-0002-1056-5895]{Weicong Jing (景威聪)}
\affiliation{National Astronomical Observatories, Chinese Academy of Sciences, Beijing 100101, China}

\author[0000-0002-9815-5873]{Jiguang Lu (卢吉光)}
\affiliation{National Astronomical Observatories, Chinese Academy of Sciences, Beijing 100101, China}
\affiliation{Guizhou Radio Astronomical Observatory, Guizhou University, Guiyang 550000, China}

\author[0009-0008-7805-091X]{Jiangwei Xu (徐江伟)}
\affiliation{National Astronomical Observatories, Chinese Academy of Sciences, Beijing 100101, China}

\author{Renxin Xu (徐仁新)}
\affiliation{Department of Astronomy, Peking University, Beijing 100871, China}
\affiliation{Kavli Institute for Astronomy and Astrophysics, Peking University, Beijing 100871, China}
\affiliation{State Key Laboratory of Nuclear Physics and Technology, Peking University, Beijing 100871, China}

\author[0009-0001-6555-3269]{Zihan Xue (薛子涵)}
\affiliation{Department of Astronomy, Peking University, Beijing 100871, China}
\affiliation{Kavli Institute for Astronomy and Astrophysics, Peking University, Beijing 100871, China}



\begin{abstract}

We use the Five-hundred-meter Aperture Spherical radio Telescope to observe the bright millisecond pulsar PSR~B1937+21 (J1939+2134) and record the data in the band from 1.0 to 1.5~GHz. We measure the neutral hydrogen (\HI) emission and absorption lines near 1420~MHz ($\lambda\simeq21\,\mathrm{cm}$). We derive the kinematic distance of the pulsar with the \HI~observation. By comparing this with the archival absorption spectra observed decades ago, we notice possible variations in the absorption spectra toward this pulsar, which correspond to a possible tiny-scale atomic structure of a few astronomical units in size. We also verify the apparent faster-than-light anomalous dispersion at the \HI~absorption line of this pulsar previously reported.

\end{abstract}



\section{Introduction}
The interaction of the magnetic moments of the electron and proton
is at the energy level of $\mathrm{\mu eV}$, which leads to the `hyperfine splitting' of the energy level between the hydrogen atom `triplet' and `singlet' states. The energy of the triplet state is higher than the singlet state by approximately $6\,\mathrm{\mu eV}$, i.e. 1420.4~MHz. \cite{1945NTvN...11..210V} predicted the 21 cm emission from \HI atoms in the space, which was later detected in 1951 \citep{1951Natur.168..356E,1951Natur.168..357M}. In the astronomical context, the \HI~absorption line was detected a couple years later \citep{1954ApJ...120..368H,1954AJ.....59Q.323H,1954Natur.173.1182W}. It was immediately proposed that the radial velocity of emission and absorption lines, combined with the Galactic rotation curve, can be used to measure the distance of radio sources \citep{1954Natur.173.1182W}, which is known as the \HI~kinematic distance. \cite{1974A&A....32..441G} and \cite{1974A&A....37..405G} first applied this method to the determination of pulsar distance. This method measures the distances of the emitting \HI~clouds in the background and the absorbing ones in the foreground, thus constraining the distance of a pulsar between the nearest background and farthest foreground clouds. Several surveys using this method have been published (see \citealt{2012ApJ...755...39V} for a summary).

Beyond distance measurement, \HI~observations of pulsars help to resolve interstellar medium (ISM) fluctuations. The angular fluctuations in the \HI~absorption spectra were firstly discovered in the very long baseline interferometry (VLBI) observation of the extragalactic source 3C~147 \citep{1976ApJ...206L.113D}, which reflects tiny-scale atomic structure (TSAS; \citealt{2018ARA&A..56..489S}) in the neutral hydrogen atoms (\HI) of the ISM. 
The temporal variations in the \HI~absorption spectra toward pulsars can also be used to detect TSAS, which was noticed by \citealt{1988ApJ...333..332C} in the absorption spectra of PSR~B1821+05. Several surveys were dedicated to the temporal variation of the \HI~absorption line toward pulsars using the Arecibo Telescope \citep{1994ApJ...436..144F,2003ApJ...598L..23S,2007ASPC..365...28W,2010ApJ...720..415S}, the Murriyang Telescope at Parkes Observatory \citep{2003MNRAS.341..941J}, and the Green Bank Telescope (GBT; \citealt{2005ApJ...631..376M}).

The \HI~energy level also affects the timing of pulsar radio signal around the transition frequency due to the electromagnetic wave propagation, as inferred from the Kramers--Kronig relation \citep[e.g.,][]{1998clel.book.....J}. The group velocity of electromagnetic wave propagating in the medium is
\begin{equation}
v_\mathrm{g}=\frac{\mathrm d\omega}{\mathrm dk}=\frac{c}{n(\omega)+\omega\frac{\mathrm d n}{\mathrm d\omega}},
\end{equation}
where $\omega$ is the angular frequency of the wave, $n=ck/\omega$ is the refractive index of the medium, and $c$ is the speed of light in vacuum \citep{1998clel.book.....J}.
In most cases, $v_\mathrm{g}$ is slower than $c$. However, around absorption lines, the group velocity can be faster than $c$, if $\mathrm dn/\mathrm d\omega$ is sufficiently negative, i.e. in the strong anomalous dispersion regime \citep{1970PhRvA...1..305G,1998clel.book.....J}. It is worth noting that the superluminal propagation (group velocity) around absorption lines does not violate causality. When it happens, the leading edge of a pulse is less attenuated than the trailing edge, thus the peak which defines the group velocity moves faster than the leading edge \citep{1998clel.book.....J}. The wide-band and periodic pulses from pulsars can be used to directly measure such anomalous dispersion of electromagnetic waves in the ISM. Indeed, 15~yr ago, the pioneer work \citep{2010ApJ...710.1718J} detected such ``fast-than-light'' propagation phenomenon around the \HI~ frequency in the Arecibo data of PSR~B1937+21 (J1939+2134). To detect such a phenomenon, one requires radio telescopes of great sensitivity that can accurately measure the time of arrivals in a rather narrow bandwidth of $\sim 100\,\mathrm{kHz}$.

The Five-hundred-meter Aperture Spherical radio Telescope (FAST) concluded its commissioning and started science observation in 2020 \citep{2019SCPMA..6259502J,2020RAA....20...64J,2020Innov...100053Q}. As the largest and most sensitive radio telescope in the L band (frequency around 1.4~GHz), FAST is capable of measuring the \HI~absorption line and anomalous dispersion accurately. In fact, such an experiment was proposed during the commissioning \citep{2020SCPMA..6329531L}. Recently, \cite{2023MNRAS.523.4949J} constrained the \HI~kinematic distance of PSR~B0458+46 using the spectral line backend of FAST.

In this paper, we focus on PSR~B1937+21 (J1939+2134), the first millisecond pulsar ever discovered \citep{1982Natur.300..615B}. Shortly after the discovery, \citet{1983ApJ...273L..75H} obtained its \HI~absorption spectrum and kinematic distance using the Arecibo Telescope. Though the kinematic distance is often less accurate compared with annual parallax measurement using pulsar timing or VLBI, it can estimate the distance of a pulsar in a single observation. The most recent astrometric result of PSR~B1937+21 derives parallax distance $2.9_{-0.2}^{+0.3}\,\mathrm{kpc}$ \citep{2023MNRAS.519.4982D}. The dispersion measures (DMs) of pulsars can also be used to estimate their distances. With $\mathrm{DM}=71\,\mathrm{cm^{-3}\,pc}$ for PSR~B1937+21, the NE2001 model \citep{2002astro.ph..7156C,2024RNAAS...8...17O} estimates its distance at 3.6~kpc, while the estimation of the YMW16 model \citep{2017ApJ...835...29Y} is 2.9~kpc. We repeat the measurement of anomalous dispersion at \HI~absorption line by \citet{2010ApJ...710.1718J}. We also notice the variations in the \HI~absorption spectrum of PSR~B1937+21 in Arecibo and FAST observations, which may be caused by a possible TSAS.

Section~\ref{sec:obs} describes the baseband observation of PSR~B1937+21 using FAST. We obtain the \HI~emission and absorption spectra in Section~\ref{sec:method_fold}, and decompose the spectra into Gaussian components in Section~\ref{sec:method_decompose} to derive the kinematic distance of PSR~B1937+21 in Section~\ref{sec:method_distance}. The results are presented in Section~\ref{sec:result}. We compare the results with previous literature in Section~\ref{sec:discussion}, and the possible TSAS is discussed in Section~\ref{sec:discussion_tsas}.

\section{Observation}\label{sec:obs}
The baseband data were recorded during a FAST observation of PSR~B1937+21 (J1939+2134) on 2020 November 12 (MJD 59165). The observation was carried out within the small zenith angle range of FAST ($<26.^\circ4$) to optimize sensitivity and polarimetry accuracy. The observation length was 1 hr. We used the central beam of the L band 19-beam receiver of FAST to track the pulsar at $\mathrm{RA}=19^\mathrm{h}39^\mathrm{m}38^\mathrm{s}.56$, $\mathrm{Dec}=+21^\circ 34'59.1''$ (J2000), or Galactic coordinates $l=57^\circ 30'32.0''$ and $b=-0^\circ 17'22.5''$. The half-power beam width (HPBW) is $2.82'$ at 1420~MHz \citep{2020RAA....20...64J}. The baseband signals of dual polarizations were sampled and recorded with a \textsc{ROACH}-2 based digital backend \citep{2019SCPMA..6259502J}, which sampled at the rate of $10^9$ samples per second. The recorded signal covered 500-MHz observing bandwidth according to Nyquist sampling theorem, i.e. 1 -- 1.5~GHz. The backend also simultaneously recorded filterbank data in the search-mode \texttt{PSRFITS} format \citep{2004PASA...21..302H} with 4 coherency matrix elements (AABBCRCI), 4096 frequency channels, and a sampling time of $49.152\,\mathrm{\mu s}$. The gain difference between the two polarizations is calibrated with the modulated noise calibrator signal from the noise diode. The noise signal was injected for 2 minutes before and after the observation.

\section{Data Reduction and Analysis}
\subsection{Channelization, Folding and Calibration}\label{sec:method_fold}
We use \textsc{dspsr} \citep{2011PASA...28....1V} to channelize and fold the baseband data off-line. We divide the 500~MHz bandwidth into 32,768 frequency channels, therefore the frequency resolution is $15.26\,\mathrm{kHz/chan}$ and the Nyquist time resolution is $65.536\,\mathrm{\mu s}$. The dynamic spectrum is folded and interpolated into 1024 phase bins, in which the spin period in the pulsar ephemeris is updated by timing segments of our observation using \textsc{tempo2} \citep{2006MNRAS.369..655H}. We access the folded archive file with the \textsc{python} language interface of \textsc{psrchive} \citep{2004PASA...21..302H}. Thus, the dispersion delay between channels is removed with the \textsc{psrchive}, and the dispersion within the channel is coherently dedispersed with the \textsc{dspsr}. The DM has been updated to $71.014527\pm 0.000006\,\mathrm{cm^{-3}\,pc}$ referencing to the Barycentric Coordinate Time (TCB) by aligning the short structures in giant pulses following the \texttt{DM\_phase} algorithm \citep{2018Natur.553..182M,2019ascl.soft10004S}. Polarization calibration is also performed using the \textsc{psrchive}. 

We estimate the antenna temperature by comparing the off-pulse spectral baseline with the system temperature, then the flux density is derived using the antenna gain. The system temperature and aperture efficiency are interpolated to the zenith angle in the middle of the observation using the parameters in \cite{2020RAA....20...64J}. We note that the \HI~emission in the beam may affect the measurement of the injected calibrator signal of noise diode. Therefore, the calibration parameters around the \HI emission line are interpolated using the parameters at neighboring frequencies.

\subsection{Spectral line decomposition}\label{sec:method_decompose}
The \HI~gas in the antenna beam contributes to the emission lines in both on- and off-pulse spectra, while only the \HI~gas in the foreground of the pulsar absorbs the pulsar emission at 21~cm. The \HI~spectral line is shifted in central frequency by the radial Doppler effect of the gas and broadened by the turbulent and thermal motion of the gas. By fitting the emission and absorption components one can derive the kinematic distance of the \HI~gas (Section~\ref{sec:method_distance}) and reveal the connection between absorption and anomalous dispersion (Section~\ref{sec:method_dispersion}).

\begin{figure}
    \centering
    \includegraphics[width=3in]{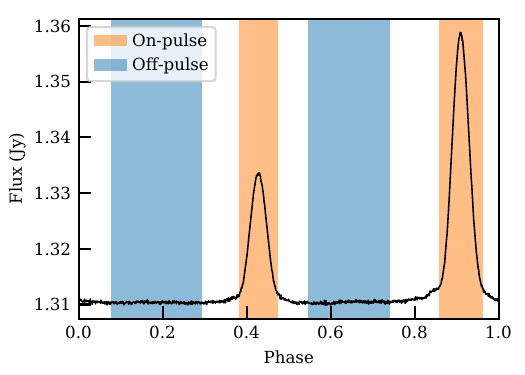}
    \caption{Integrated pulsar profile of PSR~B1937+21. The baseband signal at 1 -- 1.5~GHz is filtered into 32,768 frequency channels and folded into 1024 phase bins. The on-pulse (orange) and off-pulse (blue) ranges are visually selected.}
    \label{fig:onoff}
\end{figure}

The selected on- and off-pulse phase ranges are illustrated Fig.~\ref{fig:onoff}, where the phase ranges with less than 1\% and above 3\% peak flux are manually selected as the off- and on-pulse regions, respectively. The off-pulse spectrum, which is derived by averaging over the off-pulse phases, consists of the system noise and \HI~emission line. The on-pulse spectrum consists of the system noise, \HI~emission line, and pulsar signal with \HI~absorption. Therefore, we compute the pulsar spectrum with \HI~absorption with the on-off spectral subtraction.

As shown in Fig.~\ref{fig:kinematic_distance}, there are multiple components in the emission and absorption spectra. In order to measure the radial velocity center and scattering of each \HI~cloud (clumps of \HI~gas), it is necessary to decompose the spectra. Using the $\chi^2$ test, we note that six Gaussian components, together with a constant spectrum baseline, are enough to model the observed spectrum. Our fitting model is
\begin{equation}
T_\mathrm{A}(f)=A_0+\sum_{i=1}^{6}A_i e^{-\frac{(f-f_{c,i})^2}{2f_{d,i}^2}},\label{eq:emission_model}
\end{equation}
where $A_0$ is the constant baseline, and $A_i$, $f_{c,i}$ and $f_{d,i}$ are the amplitude, center, and half-width of the $i$ th Gaussian components, respectively.

The observed pulsar spectrum is affected by ISM scintillation. As an example, the Arecibo observation of PSR~B0540+23 and B2016+28 showed significant scintillation ripples in the pulsar spectra \citep{2010ApJ...720..415S}. In our observation, the broadband pulsar spectrum is bright around 1420~MHz and fades as the frequency increases. Thus, the \HI~absorption lines occur on the slope of the pulsar spectrum. The scintillation ripples should be carefully removed before measuring absorptive components, otherwise they may generate false structures in the absorption spectra. We use the 1D spline with explicit internal knots \texttt{LSQUnivariateSpline} in \texttt{SciPy} \citep{2020NatMe..17..261V} to fit for the baseline and interpolate around the absorption lines. For the absorption spectrum, four Gaussian components are enough to fit the optical depth, i.e. the optical depth is
\begin{equation}
S/S_b=\exp\left[-\sum_{i=1}^{4}A_i \exp\left(-\frac{(f-f_{c,i})^2}{2f_{d,i}^2}\right)\right],\label{eq:absorption_model}
\end{equation}
where $S$ is the pulsar spectrum and $S_b$ is the baseline.

All the spectral model parameters are inferred with the standard curve-fitting Bayesian method. The sampling of posterior is performed using \texttt{multiNest} \citep{2008MNRAS.384..449F,2009MNRAS.398.1601F} and \texttt{pymultiNest} \citep{2016ascl.soft06005B}. We adopt maximum posterior estimator as the central value of our models.

\subsection{\HI~Kinematic Distance of the Pulsar}\label{sec:method_distance}
The kinematic distance is determined based on the radial velocity of the source and the general revolution of the Milky Way. If the source follows the circular orbit of the Galaxy in a certain direction, its radial velocity will depend on its distance according to the Galactic rotation curve. Therefore, by measuring the radial velocity of the source, it is possible to determine its distance. Following \cite{1989ApJ...342..272F}, the observed radial velocity of the source should be
\begin{equation}
v_r=R_0(\omega-\omega_0)\sin l\cos b,\label{eq:vr}
\end{equation}
where $R_0$ is the solar distance to the Galactic center, $l$ and $b$ are the Galactic coordinates of the source, and $\omega$ and $\omega_0$ are the angular velocity of the source and the solar system, respectively. Eq.~\eqref{eq:vr} is related to the distances to the Galactic center of the source $R$ and of the Sun $R_0$ by
\begin{equation}
\Theta_0=\omega R=\omega_0 R_0,\label{eq:omega}
\end{equation}
and is related to the distance $d$ between them by
\begin{equation}
R^2=R_0^2+d^2-2R_0d\cos l,\label{eq:cos}
\end{equation}
where $\Theta_0$ is the circular rotation speed of the Milky Way. Combining Equations~\eqref{eq:vr}--\eqref{eq:cos}, the kinematic distance of the source is\footnote{Eq.~(2) in \cite{2008ApJ...674..286W} was mistyped.}
\begin{equation}
d=R_0\left[\cos l\pm\sqrt{\left(1 + \frac{v_r}{\Theta_0 \sin l\cos b}\right)^{-2}-\sin^2l}\right].\label{eq:kinematic_distance}
\end{equation}
For the outer Galaxy ($\cos l<0$), it is easy to exclude a negative root. However, for the inner Galaxy ($\cos l>0$), the near and far roots on either sides of the tangent point are both possible, which is known as kinematic distance ambiguity \citep{2012MNRAS.420.1656U}.

When applying this method to pulsars, the procedure is complicated due to lack of spectral lines in the pulsar radiation. The observer can only measure the spectral lines of foreground and background clouds to constrain the pulsar distance. For foreground clouds, the observer detects the emission line from the clouds and the absorption line in the pulsar emission. For background clouds, only emission lines can be detected. Thus the pulsar distance is constrained between the farthest absorptive cloud and the nearest emission-only cloud.

In this work, we adopt the distance to the Galactic center $R_0=8.34\,\mathrm{kpc}$ and the circular rotation speed $\Theta_0=240\,\mathrm{km\,s^{-1}}$ at the Sun \citep{2014ApJ...783..130R}. The rotation speed is measured in the local standard of rest (LSR), in which the barycentric velocity is $(U, V, W)_\odot = (11.1, 12.24, 7.25)\,\mathrm{km\,s^{-1}}$ defined in Galactic Cartesian velocity components \citep{2010MNRAS.403.1829S}. These updated values are slightly different from the current International Astronomical Union (IAU) standard $\Theta_0=220\,\mathrm{km\,s^{-1}}$, $R_0=8.5\,\mathrm{kpc}$, and $(U, V, W)_\odot = (10.0, 15.4, 7.8)\,\mathrm{km\,s^{-1}}$ \citep{1986MNRAS.221.1023K}. The choice of the above galactic model parameters introduce negligible effects compared to the systematics of kinematic distance method (about $2\,km\,s^{-1}$ in radial velocity and about 1~kpc or less in kinematic distances).

The topocentric frequency of the \HI~spectrum is first corrected to the barycenter of the solar system. Then the barycentric radial velocity of \HI~emission and absorption lines are derived from their frequency difference from the reference frequency 1420.4057517667~MHz measured in the laboratory \citep{2006sham.book.....D}. The radial velocity in the LSR, $v_\mathrm{LSR}$, is then obtained by adding the projection toward the source of $(U, V, W)_\odot$. An uncertainty of $\pm 7\,\mathrm{km\,s^{-1}}$ is added to the radial velocity due to the streaming and random motion of the clouds in the Galaxy \citep{1990AJ....100..743F,2008ApJ...674..286W}. The kinematic distance of the pulsar is derived from the off-pulse emission spectrum and the on-pulse absorption spectrum. The pulsar distance must be larger than all absorption components' distances in the on-pulse absorption spectrum, and probably smaller than the distances of the components that only exist in the off-pulse emission spectrum.

\subsection{Timing and the anomalous dispersion}\label{sec:method_dispersion}
We use pulsar-timing techniques to measure the pulse time of arrival at each frequency \citep{lorimer2005handbook}.
The timing template is averaged between 1415 and 1425~MHz and smoothed, with the frequencies of the \HI~line excluded. We implement the Fourier phase gradient algorithm \citep{1992RSPTA.341..117T} to time the integrated pulse profiles around the \HI~absorption line with a \texttt{Python} script.

The timing results are used as independent data to validate our absorption spectrum measurement. As mentioned in Section~\ref{sec:method_decompose}, ISM scintillation may generate ripples in the absorption spectra. In addition, \citet{1980A&A....88...84W} and \citet{2010ApJ...720..415S} pointed out that the small number of voltage levels in the digital spectrometer can generate large digitization errors and ``ghost'' in the spectrum. The noise temperature may also increase significantly due to bright \HI~emission, which decreases the signal-to-noise ratio (S/N) in the absorption spectrum \citep{2003MNRAS.341..941J}. While the spectral measurement is vulnerable to artifacts due to ISM and instrumental effects, the pulsar timing is less affected. According to the Kramers--Kronig relations (causality of the Green's function), dispersion must arise when resonance radiation or absorption occurs \citep{1998clel.book.....J}. The Lorentz dispersion model is a classical model that considers electrons in the medium as the damped harmonic oscillators. We follow \cite{2010ApJ...710.1718J} to assume that the radial velocity distribution of each \HI~cloud is Gaussian, i.e. the \HI~optical depth $\tau(f)=\tau_0\exp\left[-\frac{(f-f_c)^2}{2f_d^2}\right]$. In Eq.~(16) of \citet{2010ApJ...710.1718J}, the dispersion delay $\Delta$ is related to the optical depth amplitude $\tau_0$ and line width $f_d$ in the absorption profile by\footnote{A factor $\sqrt{2}$ is missing in Eq.~(16) in \cite{2010ApJ...710.1718J}.}
\begin{equation}
\Delta (f) = \frac{\tau_0}{2\sqrt{2} \pi f_d}\left[\frac{f-f_c}{\sqrt{2}f_d}\mathrm{Im}\,w\left(\frac{f-f_c}{\sqrt{2}f_d}\right)-\frac{1}{\sqrt{\pi}}\right] = \frac{\tau_0}{2\sqrt{2} \pi f_d}\left[\frac{f-f_c}{\sqrt{2}f_d}\exp\left({-\frac{(f-f_c)^2}{2f_d^2}}\right)\mathrm{erfi}\left(\frac{f-f_c}{\sqrt{2}f_d}\right)-\frac{1}{\sqrt{\pi}}\right].\label{eq:link}
\end{equation}
where the Faddeeva function $w(z)=\exp(-z^2)\mathrm{erfc}(-iz)=\frac{i}{\pi}\int_{-\infty}^{+\infty}\frac{e^{-t^2}}{z-t}\mathrm dt$ \citep{1972hmfw.book.....A}, and the imaginary error function $\mathrm{erfi}(x)=\frac{2}{\sqrt{\pi}}\int_0^xe^{t^2}\mathrm dt$. $\Delta(f)$ reaches first at the line center where $\Delta_c=-\frac{\tau_0}{(2\pi)^{3/2}f_d}$, which depends on the column density of the \HI~cloud and the velocity dispersion inside it. According to Eq.~\eqref{eq:link}, the anomalous dispersion is sensitive to narrow and deep absorptive features in the spectrum. In this way, we can infer the absorption spectra from the dispersion delay. 

\section{Results}\label{sec:result}
\subsection{Spectral Line Decomposition and Kinematic Distance}\label{sec:result_distance}
\begin{figure}[!hbt]
    \centering
    \includegraphics[width=3.5in]{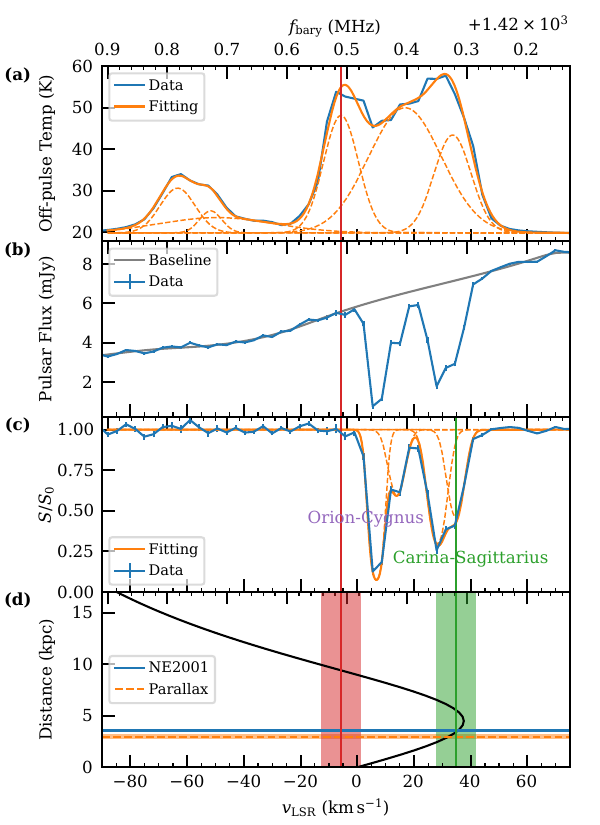}
    \caption{The kinematic distance of PSR~B1937+21. The upper $x$-axis is the barycentric frequency of the spectra. The lower $x$-axis is the radial velocity in the local standard of rest (LSR). (a) The emission spectrum of antenna temperature at off-pulse phases. The observed spectrum (blue) is fitted with 6 Gaussian components and a constant baseline. The components are plotted with orange dashed curves, and their summation is plotted with the orange solid curve.  (b) The absorption spectrum of pulsar emission. The observed spectrum is marked blue. The gray curve shows the fitted baseline. (c) The absorbed fraction. The blue curve is the observed absorption spectrum divided by the baseline in panel (b). The fitted Gaussian components are plotted with orange dashed lines, and the total absorption is plotted with the orange solid line. (d) The radial velocity of Galactic rotation as a function of distance at the direction of PSR~B1937+21.  The vertical red line denotes the radial velocity of the fourth Gaussian component in the emission spectrum and the red shade denotes its $\pm 7\,\mathrm{km\,m^{-1}}$ error, which determines the upper bound of the kinematic distance in the Perseus Arm. The vertical green line denotes the radial velocity of the fourth Gaussian component in the absorption spectrum and the green shade denotes its $\pm 7\,\mathrm{km\,m^{-1}}$ error, which determines the lower bound of the kinematic distance in the Carina--Sagittarius Arm. The horizontal blue line is the DM distance estimated using the YMW16 model \citep{2017ApJ...835...29Y}. The horizontal orange dashed line is the distance calculated from the VLBI parallax \citep{2023MNRAS.519.4982D}.}
    \label{fig:kinematic_distance}
\end{figure}

\begin{figure}[!hbt]
    \centering
    \includegraphics[width=3in]{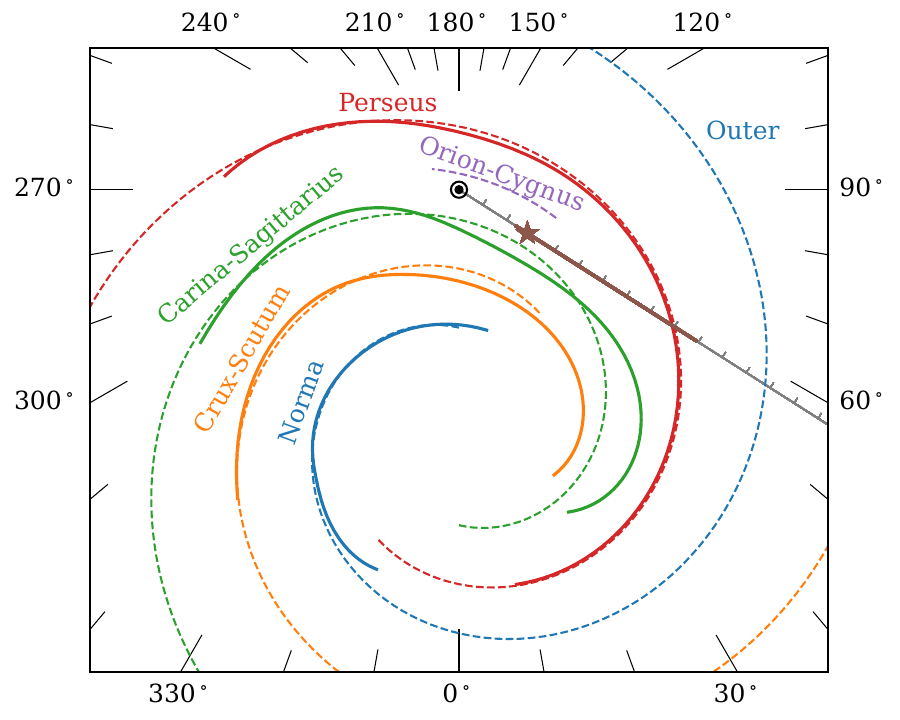}
    \caption{Distance measurement of PSR~B1937+21 in the Milky Way. The galactic arms are denoted with different colors, viz. Norma--Outer Arm (blue), Crux--Scutum Arm (orange), Carina--Sagittarius Arm (green), Perseus Arm (red), and the local Orion--Cygnus Arm (purple). The solid curves represent the Galactic arm model by \citet{1993ApJ...411..674T}, while the dashed curves represents the model by \citet{1992ApJS...83..111W}. The circle-dot represents the sun. The gray dashed line shows the line of sight toward PSR~B1937+21, with knots separated by 1~kpc. The brown solid line denotes the kinematic distance of the pulsar derived in this article, and the brown star represents its parallax distance $2.9_{-0.2}^{+0.3}\,\mathrm{kpc}$ \citep{2023MNRAS.519.4982D}.}
    \label{fig:spiral}
\end{figure}

The off-pulse emission spectrum is decomposed into six Gaussian components and a constant baseline as shown in the panel (a) of Fig.~\ref{fig:kinematic_distance}. The fitted parameters of the components are presented in Table~\ref{tab:fit_emission}. According to Figures~\ref{fig:kinematic_distance} and \ref{fig:spiral}, and Fig.~3 of \citet{2019ApJ...885..131R}, we attribute three emission components (Gaussian 4, 5, and 6 in Table~\ref{tab:fit_emission}) around $v_\mathrm{LSR}=-60\,\mathrm{km\,s^{-1}}$ to the Outer Arm of the Milky Way, the component around $-5\,\mathrm{km\,s^{-1}}$ (Gaussian 3 in Table~\ref{tab:fit_emission}) to the Perseus Arm, and the component around $35\,\mathrm{km\,s^{-1}}$ (Gaussian 1 in Table~\ref{tab:fit_emission}) to the Carina--Sagittarius Arm. The wide Gaussian component 2 in Table~\ref{tab:fit_emission} may consist of the contribution from the local Orion--Cygnus Arm, Carina--Sagittarius Arm, and Perseus Arm.
\begin{table}[!hbt]
\centering
\caption{Decomposition of the emission lines. The fitting model is defined in Eq.~\eqref{eq:emission_model}. The frequencies are in the LSR.}
\label{tab:fit_emission}
\begin{tabular}{l l l l l l}
\hline
\hline
Component & $A$ (K) & $f_c$ (MHz) & $f_d$ (MHz) & $v_\mathrm{LSR}$ ($\mathrm{km\,s^{-1}}$) & Galactic Arm\\
\hline
Gaussian 1 & $23.5 \,^{+6.0}_{-0.6}$ & $1420.3234 \,^{+0.0050}_{-0.0008}$ & $0.03096 \,^{+0.00400}_{-0.00014}$ & $33.62 \,^{+0.17}_{-1.00}$ & Carina--Sagittarius\\
Gaussian 2 & $30.10 \,^{+0.26}_{-2.20}$ & $1420.4028 \,^{+0.0120}_{-0.0034}$ & $0.0611 \,^{+0.0015}_{-0.0090}$ & $16.9 \,^{+0.7}_{-2.5}$ & local/Carina--Sagittarius/Perseus\\
Gaussian 3 & $28.3 \,^{+2.4}_{-1.7}$ & $1420.5099 \,^{+0.0007}_{-0.0030}$ & $0.0291 \,^{+0.0022}_{-0.0010}$ & $-5.72 \,^{+0.60}_{-0.15}$ & Perseus\\
Gaussian 4 & $3.7 \pm0.7$ & $1420.717 \,^{+0.013}_{-0.017}$ & $0.095 \,^{+0.020}_{-0.006}$ & $-49.4 \,^{+4.0}_{-2.8}$ & Outer\\
Gaussian 5 & $5.3 \,^{+1.6}_{-0.9}$ & $1420.728 \,^{+0.006}_{-0.005}$ & $0.0186 \,^{+0.0050}_{-0.0030}$ & $-51.7 \,^{+1.0}_{-1.2}$ & Outer\\
Gaussian 6 & $10.8 \pm0.8$ & $1420.783 \pm0.004$ & $0.0294 \,^{+0.0025}_{-0.0040}$ & $-63.3 \pm0.9$ & Outer\\
Baseline & $19.865 \,^{+0.031}_{-0.100}$\\
\hline
\end{tabular}
\end{table}

The baseline in the pulsar spectrum is fitted as shown in panel (b) of Fig.~\ref{fig:kinematic_distance}. After removing the baseline due to ISM scintillation, we identify four absorption components in the spectrum as shown in panel (c) of Fig.~\ref{fig:kinematic_distance}. We attribute Gaussian components 3 and 4 to the local Orion--Cygnus Arm, and Gaussian components 1 and 2 to Carina--Sagittarius Arm.
\begin{table}[!hbt]
\centering
\caption{Decomposition of the absorption lines. The fitting model is defined in Eq.~\eqref{eq:absorption_model}. The frequencies are in the LSR.}
\label{tab:fit_absorption}
\begin{tabular}{l l l l l l}
\hline
\hline
Component & $A$ & $f_c$ (MHz) & $f_d$ (MHz) & $v_\mathrm{LSR}$ ($\mathrm{km\,s^{-1}}$) & Galactic Arm\\
\hline
Gaussian 1 & $0.77 \,^{+0.09}_{-0.08}$ & $1420.3182 \,^{+0.0034}_{-0.0025}$ & $0.0127 \,^{+0.0021}_{-0.0015}$ & $34.7 \,^{+0.5}_{-0.7}$ & Carina--Sagittarius\\
Gaussian 2 & $1.21 \,^{+0.09}_{-0.11}$ & $1420.3488 \,^{+0.0021}_{-0.0017}$ & $0.0132 \,^{+0.0013}_{-0.0015}$ & $28.3 \,^{+0.4}_{-0.5}$ & Carina--Sagittarius\\
Gaussian 3 & $0.52 \pm0.04$ & $1420.4167 \,^{+0.0014}_{-0.0013}$ & $0.0125 \,^{+0.0017}_{-0.0013}$ & $13.94 \,^{+0.26}_{-0.30}$ & local\\
Gaussian 4 & $2.60 \pm0.17$ & $1420.4508 \pm0.0006$ & $0.0093 \pm0.0004$ & $6.75 \pm0.12$ & local\\
\hline
\end{tabular}
\end{table}

The lower bound of the kinematic distance is derived from Gaussian component 4 in Table~\ref{tab:fit_absorption} at $v_\mathrm{LSR}\approx 35\,\mathrm{km\,s^{-1}}$. This component is close to the tangent point in Carina--Sagittarius Arm, which is within the $7\,\mathrm{km\,s^{-1}}$ error range due to the streaming and random motion of the clouds. Therefore, we adopt the both larger and smaller solutions in Eq.~\eqref{eq:kinematic_distance} for a $7\,\mathrm{km\,s^{-1}}$ error as the uncertainty of the lower bound of kinematic distance, which is determined at $D_L=4.5\pm 2.1\,\mathrm{kpc}$.

The upper bound of the kinematic distance is derived from Gaussian component 4 in Table~\ref{tab:fit_emission} at $v_\mathrm{LSR}\approx -5\,\mathrm{km\,s^{-1}}$, where no absorption line can be identified. Considering a $7\,\mathrm{km\,s^{-1}}$ error due to the streaming and random motion of the clouds, the upper bound of the kinematic distance is $D_U=9.4\pm 0.5\,\mathrm{kpc}$ according to Eq.~\eqref{eq:kinematic_distance}.

\subsection{Anomalous dispersion}\label{sec:result_dispersion}
\begin{figure}[!hbt]
\centering
\includegraphics[width=3in]{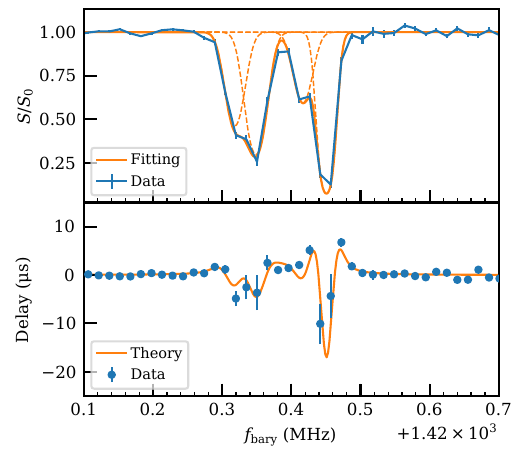}
\caption{Anomalous dispersion around the \HI~absorption line in the emission of PSR~B1937+21. The upper panel is identical to panel (c) in Fig.~\ref{fig:kinematic_distance}. The lower panel shows the dispersion delay of the pulses, in which the blue dots are measured TOAs, and the orange curve is the theoretical result derived from the absorption spectrum using Eq.~\ref{eq:link}.}
\label{fig:result_dispersion}
\end{figure}

The lower panel in Fig.~\ref{fig:result_dispersion} exhibits the measured times of arrival (TOAs) around the \HI~absorption line in the pulsar emission. At the center of the fourth absorption component, the dispersion delay reaches $-10\,\mathrm{\mu s}$, i.e. the pulse is faster than light by about $10\,\mathrm{\mu s}$. With the Gaussian decomposition of the absorption spectrum in the upper panel of Fig.~\ref{fig:result_dispersion}, we derive the theoretical delay curve in the lower panel from Eq.~\ref{eq:link}, which matches well with the observed TOAs.

\section{Discussion}\label{sec:discussion}
\subsection{A possible tiny-scale atomic structure}\label{sec:discussion_tsas}
PSR~B1937+21 is associated with the continuum radio source 4C~21.53W, which is a slightly extended \HII~region \citep{1984A&A...130..257S}. In the vicinity, 4C~21.53E (1938+215) is an extragalactic double source \citep{1983ApJ...264L..13E}. The supernova remnant (SNR) G57.2+0.8 (1932+218) is also known as 4C~21.53 \citep{2018OPhyJ...4....1R,2024ApJS..270...21C}, which is the host of magnetar SGR~J1935+2154 and is not related to PSR~B1937+21.

\begin{figure}[!hbt]
    \centering
    \includegraphics[width=3in]{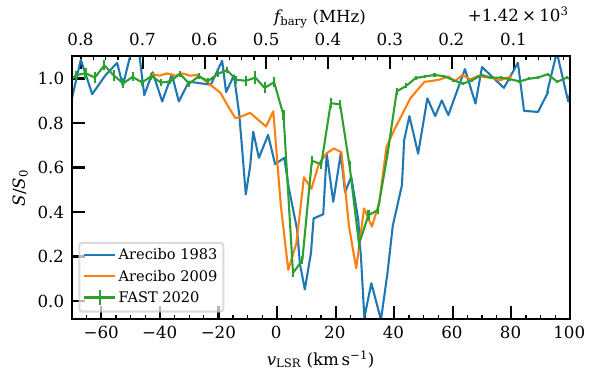}
    \caption{Temporal evolution of the \HI~absorption spectrum toward PSR~B1937+21. The blue curve is the Arecibo result extracted from Fig.~1 of \cite{1983ApJ...273L..75H}. The orange curve is the Arecibo result extracted from Fig.~3 of \citet{2010ApJ...710.1718J}. A velocity offset $-20\,\mathrm{km\,s^{-1}}$ is applied to align its corresponding emission spectrum with the FAST emission spectrum, because the frequency in \citet{2010ApJ...710.1718J} was topocentric. The data points are digitized using \texttt{PlotDigitizer} (\url{https://plotdigitizer.com/app}). The green curve is the result of the current work.}
    \label{fig:comparison}
\end{figure}

The absorption lines in the pulsar spectra of PSR~B1937+21 were previously measured by \cite{1983ApJ...273L..75H} and \cite{2010ApJ...710.1718J} using the Arecibo Telescope. Fig.~1 of \cite{1983ApJ...273L..75H} presented a narrow absorption feature at $-10\,\mathrm{km\,s^{-1}}$ in the absorption spectra toward PSR~B1937+21 and 4C~21.53W, which was suspected to be absorbed by a nearby small dust cloud. This narrow absorption line disappeared in the absorption spectrum toward this pulsar, as shown in Fig.~2 of \cite{2010ApJ...710.1718J} and in Fig.~\ref{fig:kinematic_distance} of this article. We also notice that a wider and shallower component connected the $-10\,\mathrm{km\,s^{-1}}$ component to the $+10\,\mathrm{km\,s^{-1}}$ component in the absorption spectra in \cite{1983ApJ...273L..75H} and \cite{2010ApJ...710.1718J}, which is also missing in our result. The comparison of the three absorption spectra is presented in Fig.~\ref{fig:comparison}.

If the observed variation in the \HI~absorption spectrum is caused by transverse motion of \HI~cloud across the line of sight (LoS) toward the pulsar, the transverse size of the \HI~cloud can be roughly estimated. The transverse size of the cloud can be derived from the variation time scale of \HI~absorption line and the transverse velocity of the cloud relative to the LoS. However, our knowledge of both factors is very limited. The \HI~absorption spectrum of PSR~B1937+21 was only measured in three epochs separated by decades, which can only poorly imply the variation time scale. The proper motion and distance of the cloud are also unknown. Following previous studies such as \citet{2005ApJ...631..376M}, we estimate the transverse speed by multiplying the proper motion of PSR~B1937+21 with half of its distance. We must point out the large uncertainty in this estimation. With the proper motion $0.24\,\mathrm{mas\,yr^{-1}}$ and the distance $D=1/\varpi=1/(0.34\,\mathrm{mas}) =2.9\,\mathrm{kpc}$ of PSR~B1937+21 \citep{2023MNRAS.519.4982D}, the variation between 1983 and 2009 corresponds to a spatial structure around 9~au at half way toward the pulsar, and the variation between 2009 and 2020 corresponds to a structure around 4~au, assuming the \HI~cloud is at halfway toward the pulsar.

Temporal variations of \HI~absorption profiles toward pulsars have been reported for several pulsars, including PSR~B1821+05 \citep{1988ApJ...333..332C,1991ApJ...382..168F}, B1557$-$50 \citep{1992MNRAS.258P..19D,2003MNRAS.341..941J}, and B0301+19 \citep{2008ApJ...674..286W}. The variations are attributed to the TSASs in the ISM \citep{2018ARA&A..56..489S}. The TSAS toward PSR~B1937+21 has not been reported in previous literature. This detection provides a new source to monitor the tiny-scale fluctuation of \HI~in the ISM.

However, we must emphasize that the measurement of \HI~absorption spectra is vulnerable to possible errors generated during observation and data reduction, as mentioned in Section~\ref{sec:method_dispersion}. \citet{2003MNRAS.341..941J} pointed out the lack of significance contours manifesting the noise increase due to \HI~emission in the earlier study by \citet{1994ApJ...436..144F}. \citet{1980A&A....88...84W} and \citet{2010ApJ...720..415S} noted that the limited number of voltage levels can generate large digitization errors and ``ghost'' in the spectrum. We also find that the pulsar spectrum of PSR~B1937+21 shows strong scintillation ripples, as also noticed by \citet{2010ApJ...720..415S}. Therefore, the baseline removal is crucial in the \HI~absorption line measurement.

There are also possible variations in other absorption components. In the absorption spectrum in \citet{2010ApJ...710.1718J}, the two prominent components at 1420.26 and 1420.37~MHz (topocentric frequency) showed similar amplitude. In our observation, the higher-frequency peak is deeper. The anomalous dispersion presented in Section~\ref{sec:result_dispersion} and Fig.~\ref{fig:result_dispersion} may also provide evidence for the possible TSAS. Unlike the spectral line measurement, the digitization error and ISM scintillation should affect the errors of TOAs rather than their expectations. The dispersion delay  observed in our observation (lower panel of Fig.~\ref{fig:result_dispersion}) is significantly different from Fig.~3 of \citet{2010ApJ...710.1718J}, which implies that the absorption spectrum has also changed. Unfortunately, the dispersion delay was not measured in \citet{1983ApJ...273L..75H}, and the dispersion is insensitive to the broad component around 1420.45~MHz (topocentric frequency) in \citet{2010ApJ...710.1718J}; therefore the previously discussed TSAS around $v_\mathrm{LSR}=-5\,\mathrm{km\,s^{-1}}$ cannot be fully assured using anomalous dispersion.

\subsection{Update of kinematic distance of PSR~B1937+21}
The kinematic distance of PSR~B1937+21 has been measured in several previous studies. \citet{1983ApJ...273L..75H} acquired the \HI~emission and absorption spectra of PSR~B1937+21 using the Arecibo Telescope shortly after its discovery \citep{1982Natur.300..615B}. Their pulsar spectrum did not show as much absorption around $v_r=40\,\mathrm{km\,s^{-1}}$ as the nearby \HII~region 4C~21.53W and extragalactic source 4C~21.53E. They concluded that the pulsar must be closer to us than the tangent point of the spiral arm, and therefore the pulsar distance should be smaller than 5~kpc. By contrast, \citet{1990AJ....100..743F} set the lower limit of pulsar distance $D_L=4.9\pm 1.9\,\mathrm{kpc}$ at the tangent point, and moved the upper limit to $D_U=14.8\pm 0.9\,\mathrm{kpc}$ with the \HI~emission line around $v_r=-60\,\mathrm{km\,s^{-1}}$.

In this paper, our lower bound of kinematic distance $D_L=4.5\pm 2.1\,\mathrm{kpc}$ is similar to the result of \cite{1990AJ....100..743F} because we identify the same \HI~absorption component near the tangent point. However, we shorten the upper bound from $14.8\pm 0.9\,\mathrm{kpc}$ \citep{1990AJ....100..743F} in the Outer Arm to $9.4\pm 0.5\,\mathrm{kpc}$ in the Perseus Arm, because we recognize Gaussian component 3 around $-5\,\mathrm{km\,s^{-1}}$ in the emission spectrum has no counterpart in the absorption spectrum. We agree with \cite{1983ApJ...273L..75H} and \citet{1990AJ....100..743F} that the absorption at approximately $-5$ and $-10\,\mathrm{km\,s^{-1}}$ was caused by local cloud with noncircular motion. This argument can be supported by DM and parallax distances. Our result is consistent with the more accurate parallax distance $2.9_{-0.2}^{+0.3}\,\mathrm{kpc}$ \citep{2023MNRAS.519.4982D}.

\subsection{Anomalous dispersion measurement in pulsar observation}
Our result confirms the pioneer discovery of apparent faster-than-light dispersion around the \HI~absorption line \citep{2010ApJ...710.1718J}, though the delay curve has changed since 2009 as discussed in Section~\ref{sec:discussion_tsas}. In the lower panel of Fig.~\ref{fig:result_dispersion}, the negative delay at the line center implies that the group velocity $v_\mathrm{g}$ is greater than the speed of light in vacuum. As discussed in Section~\ref{sec:method_dispersion}, the superluminal group velocity at the center of absorption line can be derived from classical electrodynamics. However, it does not violate causality. The group velocity $v_\mathrm{g}=\mathrm d\omega/\mathrm dk$ is defined as the speed of the Gaussian peak of the wave packet \citep{1998clel.book.....J}. When radiation or absorption occurs, $v_\mathrm{g}$ is not the equivalent to the propagation speed of information \citep{2010ApJ...710.1718J}. The connection between absorption and dispersion, known as Kramers--Kronig relation, is the result of causality in the Green's function of radio wave propagation in the medium \citep{1998clel.book.....J}.

According to Eq.~\eqref{eq:link}, the amplitude of anomalous dispersion depends on both the amplitude and the width of the absorption line. The measurement of anomalous dispersion is sensitive to deep and narrow absorption features. However, deep absorption also decreases the observed brightness, which decreases the S/N of the pulse profiles and increases the errors of TOAs. Therefore, the successful measurement of anomalous dispersion in pulsar emission requires dense \HI~clouds with small velocity dispersion, and bright pulsars in the background. PSR~B1937+21 resides near the tangent point, from where the radio wave propagates through a long path in Carina--Sagittarius Arm and the local Orion--Cygnus Arm. The high flux of PSR~B1937+21 also increases the precision of timing. We recommend bright millisecond pulsars around tangent points of spiral arms for observations in the future.

\begin{acknowledgments}
We thank the anonymous reviewer and Joel Weisberg for helpful comments. This work made use of the data from FAST (Five-hundred-meter Aperture Spherical radio Telescope) (\url{https://cstr.cn/31116.02.FAST}). FAST is a Chinese national mega-science facility, operated by National Astronomical Observatories, Chinese Academy of Sciences. J.J.C. thanks Pengfei Wang and Chao Wang at NAOC, CAS and Xun Shi at YNAO, CAS for inspiring discussions. This work is supported by the National SKA Program of China (2020SKA0120100), the National Natural Science Foundation of China (Nos. 12003047 and 12133003), and the Strategic Priority Research Program of the Chinese Academy of Sciences (No. XDB0550300). This work is also supported by the Chinese Academy of Sciences Project for Young Scientists in Basic Research, grant No. YSBR-063; National Science Foundation of China 12225303, 12421003; and Strategic Priority Research Program of the Chinese Academy of Sciences, grant No.XDA0350501.
\end{acknowledgments}

%

\vspace{5mm}
\facilities{FAST.}


\software{DSPSR \citep{2011PASA...28....1V}. PSRCHIVE \citep{2004PASA...21..302H}, TEMPO2 \citep{2006MNRAS.369..655H}, MultiNest \citep{2008MNRAS.384..449F,2009MNRAS.398.1601F}, PyMultiNest \citep{2016ascl.soft06005B}, Astropy \citep{2013A&A...558A..33A,2018AJ....156..123A}, Numpy \citep{2020Natur.585..357H}, SciPy \citep{2020NatMe..17..261V}, Matplotlib \citep{2007CSE.....9...90H}, MWPROP \citep{2024RNAAS...8...17O}, PlotDigitizer (\url{https://plotdigitizer.com/app})
          }




\bibliography{reference}{}
\bibliographystyle{aasjournal}


\end{CJK*}
\end{document}